\documentclass[dvips]{acta}
\usepackage{lscape,epsfig}
\usepackage{amssymb}
\usepackage{amsmath}
\usepackage{xspace}
\usepackage{rotating}

\newcommand{\ceight}{K2-295\xspace}
\newcommand{\celeven}{K2-237\xspace}
\newcommand{\ktwo}{{\it K2}\xspace}
\newcommand{\kep}{\textit{Kepler}\xspace}
\newcommand{\feh}{\mbox{[Fe/H]}\xspace}
\newcommand{\teff}{\mbox{$T_{\rm *, eff}$}\xspace}
\newcommand{\logg}{\mbox{$\log g_*$}\xspace}
\newcommand{\vsini}{\mbox{$v \sin i_{*}$}\xspace}
\newcommand{\kms}{\mbox{km\,s$^{-1}$}\xspace}

\newcommand{\mplanet}{\mbox{$M_{\rm p}$}\xspace}
\newcommand{\rplanet}{\mbox{$R_{\rm p}$}\xspace}
\newcommand{\mjup}{\mbox{$\mathrm{M_{\rm Jup}}$}\xspace}
\newcommand{\rjup}{\mbox{$\mathrm{R_{\rm Jup}}$}\xspace}
\newcommand{\msat}{\mbox{$\mathrm{M_{\rm Sat}}$}\xspace}
\newcommand{\rsat}{\mbox{$\mathrm{R_{\rm Sat}}$}\xspace}

\newcommand{\mstar}{\mbox{$M_{*}$}\xspace}
\newcommand{\rstar}{\mbox{$R_{*}$}\xspace}
\newcommand{\densstar}{\mbox{$\rho_*$}\xspace}
\newcommand{\densplanet}{\mbox{$\rho_{\rm p}$}\xspace}
\newcommand{\msol}{\mbox{$\mathrm{M_\odot}$}\xspace}
\newcommand{\rsol}{\mbox{$\mathrm{R_\odot}$}\xspace}
\newcommand{\ecos}{\mbox{$e \cos \omega$}\xspace} 
\newcommand{\esin}{\mbox{$e \sin \omega$}\xspace}

\newcommand{\denssol}{\mbox{$\mathrm{\rho_\odot}$}\xspace}

\SetPages{0}{0}

\SetVol{58}{2018}

\begin{document}

\begin{Titlepage}
\Title{\ceight~b and \celeven~b: two transiting hot Jupiters}

\Author{A.~M.~S.~Smith$^{1*}$, Sz.~Csizmadia$^{1}$,
D.~Gandolfi$^{2}$, 
S.~Albrecht$^{3}$,
R.~Alonso$^{4,5}$,
O.~Barrag\'{a}n$^{2}$,
J.~Cabrera$^{1}$,
W.~D.~Cochran$^{6}$,
F.~Dai$^{7}$,
H.~Deeg$^{4,5}$, 
Ph.~Eigm\"uller$^{1,8}$, 
M.~Endl$^{6}$, 
A.~Erikson$^{1}$,
M.~Fridlund$^{9,10,4}$,
A.~Fukui$^{11,4}$,
S.~Grziwa$^{12}$, 
E.~W.~Guenther$^{13}$, 
A.~P.~Hatzes$^{13}$,
D.~Hidalgo$^{4,5}$,
T.~Hirano$^{14}$, 
J.~Korth$^{12}$,
M.~Kuzuhara$^{15,16}$,
J.~Livingston$^{17}$,
N.~Narita$^{17,15,16,4}$,
D.~Nespral$^{4,5}$, 
P.~Niraula$^{18}$,
G.~Nowak$^{4,5}$, 
E.~Palle$^{4,5}$,
M.~P\"atzold$^{12}$,
C.M.~Persson$^{10}$,   
J.~Prieto-Arranz$^{4,5}$,
H.~Rauer$^{1,8,19}$,
S.~Redfield$^{18}$,
I.~Ribas$^{20,21}$,
and V.~Van~Eylen$^{9}$
.}
{
$^{1}$Institute of Planetary Research, German Aerospace Center, Rutherfordstrasse 2, 12489 Berlin, Germany.
$^{2}$Dipartimento di Fisica, Universit\'a di Torino, Via P. Giuria 1, I-10125, Torino, Italy.
$^{3}$Stellar Astrophysics Centre, Department of Physics and Astronomy, Aarhus University, Ny Munkegade 120, DK-8000 Aarhus C, Denmark.
$^{4}$Instituto de Astrof\'\i sica de Canarias (IAC), 38205 La Laguna, Tenerife, Spain.
$^{5}$Departamento de Astrof\'\i sica, Universidad de La Laguna (ULL), 38206 La Laguna, Tenerife, Spain.
$^{6}$Department of Astronomy and McDonald Observatory, University of Texas at Austin, 2515 Speedway, Stop C1400, Austin, TX 78712, USA.
$^{7}$Department of Physics and Kavli Institute for Astrophysics and Space Research, Massachusetts Institute of Technology, Cambridge, MA 02139,USA.
$^{8}$Center for Astronomy and Astrophysics, TU Berlin, Hardenbergstr. 36, 10623 Berlin, Germany.
$^{9}$Leiden Observatory, Leiden University, 2333CA Leiden, The Netherlands.
$^{10}$Chalmers University of Technology, Department of Space, Earth and Environment, Onsala Space Observatory,  SE-439 92 Onsala, Sweden.
$^{11}$Subaru Telescope Okayama Branch Office, National Astronomical Observatory of Japan, NINS, 3037-5 Honjo, Kamogata, Asakuchi, Okayama 719-0232, Japan.
$^{12}$Rheinisches Institut f\"ur Umweltforschung an der Universit\"at zu K\"oln, Aachener Strasse 209, 50931 K\"oln, Germany.
$^{13}$Th\"uringer Landessternwarte Tautenburg, Sternwarte 5, 07778 Tautenburg, Germany.
$^{14}$Department of Earth and Planetary Sciences, Tokyo Institute of Technology, 2-12-1 Ookayama, Meguro-ku, Tokyo 152-8551, Japan.
$^{15}$Astrobiology Center, NINS, 2-21-1 Osawa, Mitaka, Tokyo 181-8588, Japan.
$^{16}$National Astronomical Observatory of Japan, NINS, 2-21-1 Osawa, Mitaka, Tokyo 181-8588, Japan.
$^{17}$Department of Astronomy, The University of Tokyo, 7-3-1 Hongo, Bunkyo-ku, Tokyo 113-0033, Japan.
$^{18}$Astronomy Department and Van Vleck Observatory, Wesleyan University, Middletown, CT 06459, USA.
$^{19}$Institute of Geological Sciences, Freie Universit\"at Berlin, Malteserstr. 74-100, 12249 Berlin, Germany.
$^{20}$Institut de Ci\`{e}ncies de l'Espai (ICE, CSIC), Campus UAB, Can Magrans s/n, 08193 Bellaterra, Spain.
$^{21}$Institut d'Estudis Espacials de Catalunya (IEEC), 08034 Barcelona, Spain.\\
*e-mail:Alexis.Smith@dlr.de}

\Received{Month Day, Year}
\end{Titlepage}

\Abstract{We report the discovery from K2 of two transiting hot Jupiter systems. \ceight (observed in Campaign~8) is a K5 dwarf which hosts a planet slightly smaller than Jupiter, orbiting with a period of 4.0~d. We have made an independent discovery of \celeven~b (Campaign~11), which orbits an F9 dwarf every 2.2~d and has an inflated radius 60 -- 70 per cent larger than that of Jupiter. We use high-precision radial velocity measurements, obtained using the HARPS and FIES spectrographs, to measure the planetary masses. We find that \ceight~b has a similar mass to Saturn, while \celeven~b is a little more massive than Jupiter.
}
{planetary systems  -- planets and satellites: detection -- planets and satellites: individual: \ceight -- planets and satellites: individual: \celeven
}

\section{Introduction}

Two decades after the discovery of the first hot Jupiter, there remains much to be understood about these intrinsically rare objects (e.g. Howard et al. 2012). Open questions concern the formation and migration of hot Jupiters, as well as the nature of the mechanism responsible for their inflation.

Most well-characterised hot Jupiter systems were discovered by wide-field, ground-based surveys such as WASP (Pollacco et al. 2006) and HATNet (Bakos et al. 2002). Recently, the \ktwo mission (Howell et al. 2014) has been used to discover such systems, and can determine planetary radii to greater precision. Ground-based radial velocity (RV) observations remain crucial, not only to confirm the planetary nature of the system, but to enable a fuller characterisation by measuring the planet-to-star mass ratio. It is only by increasing the sample of hot Jupiter systems with well-measured properties that we will be able to more fully understand them. 

In particular, hot Jupiters, particularly low-density or inflated planets, are attractive targets for atmospheric characterisation (e.g. Seager \& Deming 2010; Sing et al. 2016). In addition, detections of evaporating atmospheres often come from this same sample (Lyman-alpha: Vidal-Madjar et al. 2003; H-alpha: Jensen et al. 2012; HeI: Spake et al. 2018) and represent a possible mechanism for the transformation of hot gas giants into hot rocky super-Earths (Valencia et al. 2010; Lopez, Fortney \& Miller 2012).

In this paper we report the discovery, under the auspices of the KESPRINT collaboration\footnote{http://www.kesprint.science}, of \ceight and \celeven, two transiting hot Jupiter systems observed in K2 Campaigns 8 and 11, respectively. We use radial velocity follow-up measurements to confirm the planetary nature of the systems, and to measure the planetary masses. The discovery of \celeven was recently reported by Soto et al. (2018), who measured the planet's mass using RVs from the CORALIE and HARPS instruments. Here, we report an independent discovery of the same planetary system, and confirm their conclusion that the planet is inflated. We also perform a joint analysis incorporating the radial velocity data obtained by Soto et al. (2018).

\section{Observations}
\label{sec:obs}
\subsection{\ktwo photometry}
\label{sec:obs_k2}

\ceight was observed as part of \ktwo's Campaign 8, from 2016 January 04 to 2016 March 23. \celeven was observed as part of Campaign 11, which ran from 2016 September 24 to 2016 December 07. A change in the roll attitude of the spacecraft was required part way through the observing campaign. This has the effect that the C11 data are divided into two segments, with a 76-hour gap between 2016 October 18 and 21 where no observations were made\footnote{See \ktwo Data Release Notes at https://keplerscience.arc.nasa.gov/k2-data-release-notes.html}.

We used two different detection codes to search the publicly available light curves, produced by the authors of (Vanderburg \& Johnson 2014), for periodic transit-like signals. {\sc Exotrans / Varlet} (Grziwa, P\"atzold \& Carone 2012; Grizwa \& P\"atzold 2016) and {\sc DST} (Cabrera et al. 2012) detected consistent signals for both \ceight and \celeven. \ceight undergoes transits of about 2 per cent depth, approximately every 4 days, whereas the transits of \celeven are around 1.5 per cent deep, and repeat every 2.2 days. This system was  also detected using the BLS algorithm and an optimized frequency grid, described by Ofir (2014).

We also note that \ceight was recently reported as a planetary candidate by Petigura et al. (2018), who report stellar properties for this target, determined from a Keck/HIRES spectrum using {\sc SpecMatch-emp} (Yee, Petigura \& von Braun 2017). The values reported by Petigura et al. (2018) are in good agreement with those obtained from our independent data and analysis (see Section~3.2). Basic catalogue information on \ceight and \celeven are given in Table~1.

\begin{table}
\caption{Catalogue information for \ceight and \celeven.}
\begin{center}
\begin{tabular}{lll} \hline
Parameter  & \ceight & \celeven \\ \hline
RA (J2000.0) & 01h18m26.376s & 16h55m04.534s \\
Dec (J2000.0) & $+06^{\circ}~49^\prime~00^{\prime\prime}.74 $ & $-28^{\circ}~42^\prime~38^{\prime\prime}.03 $\\
pmRA$^*$ (mas yr$^{-1}$) & $54.98 \pm 0.05$  & $ -8.57 \pm 0.10$\\
pmDec$^*$ (mas yr$^{-1}$) & $-34.96 \pm 0.04$  & $-5.56 \pm 0.05$\\
parallax$^*$ (mas) & $4.27 \pm 0.03$ & $3.15 \pm 0.07$\\
\hline
Magnitudes& & \\
$B$&15.07 $\pm$ 0.08 & 12.19 $\pm$ 0.07\\
$g\prime$&14.55 $\pm$ 0.04 & 11.83 $\pm$ 0.06\\
$V$&13.95 $\pm$ 0.04 & 11.60 $\pm$ 0.05\\
$r\prime$&13.46 $\pm$ 0.03 & 11.45 $\pm$ 0.03\\
\kep&13.54 & 11.47\\
$i\prime$&13.10 $\pm$ 0.06 & 11.31 $\pm$ 0.04\\
$J$ (2MASS)&11.81 $\pm$ 0.03 & 10.51 $\pm$ 0.02\\
$H$ (2MASS)&11.26 $\pm$ 0.02 & 10.27 $\pm$ 0.02\\
$K$ (2MASS)&11.14 $\pm$ 0.03 & 10.22 $\pm$ 0.02\\
\hline
\multicolumn{3}{l}{Additional identifiers:}\\
EPIC & 220501947 (C8) & 229426032 (C11)\\
UCAC & 485-001859 & 307-097169\\
2MASS & 01182635+0649004 & 16550453-2842380\\
\hline
\multicolumn{3}{l}{$^*$Data taken from Gaia DR2.}\\
\end{tabular}
\end{center}
\label{tab:stellar}
\end{table}

\subsection{High resolution imaging}
\label{sec:obs:ao}
We obtained high resolution/contrast images of \ceight using 
the Infrared Camera and Spectrograph (IRCS; Kobayashi et al. 2000) on Subaru with the adaptive-optics system (AO188; Hayano et al. 2010) on UT 2016 November 7. We observed the target with the $H-$band filter and fine-sampling mode (1 pix = $0.^{\prime\prime}02057$). For \ceight, both saturated (36~s) and unsaturated (4.5~s) frames were repeatedly obtained with a five-point dithering, which were used to search for faint companions and absolute flux calibration, respectively. The total scientific exposure amounted to 540~s for the saturated frames. 

We observed \celeven with the Multi-color Simultaneous Camera for studying Atmospheres of Transiting exoplanets (MuSCAT; Narita et al. 2015), mounted on the 1.88-m telescope at the Okayama Astronomical Observatory. We conducted observations on UT 2017 August 7, obtaining 30 images with the exposure time of 2.5~s in the Sloan $g\prime$, $r\prime$, and $z\prime$ bands. The pixel scale of 0.36$^{\prime\prime}$ / pixel and median seeing of 2.1$^{\prime\prime}$ allow the detection of faint objects a few arcseconds away from the target star. 

We also performed Lucky Imaging (LI) of \celeven using the FastCam camera (Oscoz et al. 2008) on the 1.55-meter Telescopio Carlos S\'anchez (TCS) at Observatorio del Teide, Tenerife. FastCam is a very low noise and fast readout speed EMCCD camera with $512 \times 512$ pixels (with a physical pixel size of 16 microns, and a FoV of $21.2"\times 21.2"$). During the night of 2017 July 19 (UT), $10,000$ individual frames of \celeven were collected in the Johnson-Cousins I-band (infrared), with an exposure time of 50 ms for each frame.

\subsection{Spectroscopic observations}
\label{sec:rvobs}

We obtained a single reconnaissance spectrum of \ceight with the Robert G. Tull coud\'e spectrograph (Tull et al. 1995) on the 2.7-m Harlan J. Smith Telescope at McDonald Observatory, Texas. The goal of the observation (before the availability of Gaia DR2) was to check that the target is not an SB2, a giant star, or a fast rotator, in which cases it is likely to be a false positive, or unamenable to radial velocity follow-up observations. The observation was conducted on 2016 October 13, and the exposure time was 1611~s, yielding S/N = 35 per resolution element at 565 nm.

Radial velocity (RV) observations were performed using the FIbre-fed \'Echelle Spectrograph (FIES; Frandsen \& Lindberg 1999; Telting et al. 2014) mounted at the 2.56-m Nordic Optical Telescope (NOT) of Roque de los Muchachos Observatory (La Palma, Spain). We employed the \emph{med-res} fibre for \ceight and the \emph{high-res} fibre for \celeven, resulting in resolving powers, $R\,=\,\lambda/\Delta\lambda\,\approx\,47\,000$ and $67\,000$, respectively. We took three consecutive exposures of 900-1200 sec per observation epoch to remove cosmic ray hits. We traced the intra-exposure RV drift of the instrument by acquiring long-exposed ($\sim$40 sec) ThAr spectra immediately before and after the target observations (Gandolfi et al. 2015). The data was reduced using standard IRAF and IDL routines, which include bias subtraction, flat fielding, order tracing and extraction, and wavelength calibration. The RV measurements of \ceight and \celeven were extracted via multi-order cross-correlations with a FIES spectrum of the RV standard stars HD\,190007 and HD\,168009, respectively. Seven measurements of \ceight were secured between October 2006 and January 2017 under the observing programs 54-027 and 54-205. Ten FIES spectra of \celeven were gathered between July and August 2017 as part of the observing programs 55-019 and OPTICON 17A/064.

Additionally, we acquired seven high-resolution spectra (R$\approx$115\,000) of K2-237 with the HARPS spectrograph (Mayor et al. 2003) based on the ESO 3.6-m telescope at La Silla Observatory (Chile). The observations were performed in August 2017 as part of the ESO programme 099.C-0491. We set the exposure time to 900--1800 seconds and used the second fibre to monitor the sky background. We reduced the data with the on-line HARPS pipeline and extracted the RVs by cross-correlating the HARPS spectra with a G2 numerical mask (Baranne et al. 1996; Pepe et al. 2002).

All of our RV measurements are listed in Table~2 along with their $1\sigma$ uncertainties and the bisector spans of the cross-correlation functions.

\begin{table}
\caption{Radial velocity measurements. Entries in italics were taken during transit, and excluded from the modelling (see Section~5.3). }
\begin{center}
\begin{tabular}{lllrrc}\hline
$\mathrm{BJD_{TDB}}$ & RV & $\sigma_{\mathrm{RV}}$ & BIS & $\sigma_{\mathrm{BIS}}$ & Inst.$^\dagger$ \\
 $-2450000$ & \kms & \kms & \kms & \kms &\\
\hline\\
\ceight &&&\\
\hline
7668.668055 & -16.569 & 0.017 & 0.028 & 0.034 & F \\
7669.555227 & -16.657 & 0.012 & 0.035 & 0.024 & F \\
7682.549665 & -16.656 & 0.018 & 0.052 & 0.036 & F \\
7684.536591 & -16.576 & 0.019 & 0.034 & 0.038 & F \\
{\it 7717.374193} & {\it -16.610} & {\it 0.012} & {\it 0.040} & {\it 0.024} &{\it  F} \\
7769.395395 & -16.612 & 0.020 & 0.025 & 0.040 & F \\
7777.384377 & -16.584 & 0.018 & 0.022 & 0.036 & F \\
\hline
\celeven &&&\\
\hline
7954.463961 & -22.354 & 0.027 &  0.022 & 0.054 & F \\
7955.456791 & -22.641 & 0.061 & -0.086 & 0.122 & F \\
7956.432724 & -22.463 & 0.053 &  0.040 & 0.106 & F \\
7964.393400 & -22.707 & 0.067 &  0.086 & 0.134 & F \\
7965.402076 & -22.354 & 0.045 &  0.003 & 0.090 & F \\
7966.393358 & -22.625 & 0.042 & -0.088 & 0.084 & F \\
7980.391270 & -22.496 & 0.068 &  0.012 & 0.136 & F \\
{\it 7981.389387} & {\it -22.562} & {\it 0.050} & {\it -0.022} & {\it 0.100} & {\it F} \\
7982.387408 & -22.505 & 0.058 &  0.025 & 0.116 & F \\ 
7983.391266 & -22.353 & 0.066 &  0.004 & 0.132 & F \\
7984.556027 & -22.362 & 0.011 & -0.019 & 0.022 & H \\
7985.483048 & -22.213 & 0.016 & -0.046 & 0.032 & H \\
7986.559244 & -22.433 & 0.022 &  0.056 & 0.044 & H \\
7987.509317 & -22.164 & 0.015 & -0.038 & 0.030 & H \\
7990.472080 & -22.480 & 0.015 &  0.098 & 0.030 & H \\
7991.487944 & -22.208 & 0.012 & -0.083 & 0.024 & H \\
7992.484469 & -22.426 & 0.009 & -0.029 & 0.018 & H \\
\hline
\multicolumn{4}{c}{}$^\dagger$ F = FIES, H = HARPS\\
\\
\end{tabular}
\end{center}
\label{tab:rv}
\end{table}

\section{Stellar characterisation}
\label{sec:star}

\begin{table}
\caption{Adopted stellar parameters. See Sections 3.2 and 3.3 for a full discussion of how these values were derived.}
\begin{center}
\begin{tabular}{lcc} \hline
Parameter  &  		\ceight				& \celeven 	\\ \hline
\teff /K &			$ 4444 \pm 70 $		& $6099 \pm 110$	\\
\rstar / \rsol &	$0.70 \pm 0.02$			& $1.38 \pm 0.04$	\\
\feh (dex) &  		$0.14 \pm 0.12$		& $0.00 \pm 0.08$	\\
\mstar / \msol  & 	$0.74 \pm 0.04$		& $1.23 \pm 0.05$	\\
\vsini / \kms &		$2.2 \pm 0.3$			& $12 \pm 1$		\\
 \logg [cgs] & 	$4.63 \pm 0.12$		& $4.27 \pm 0.12$	\\
Distance / pc  & 	$225.7 \pm 2.4$		 	& $309.5 \pm 7.4$	\\
Spectral type & K5\,V						&  F9\,V	\\
\hline
\end{tabular}
\end{center}
\label{tab:spec}
\end{table}

\subsection{Method}
\label{sec:spec_method}

We adopt the following procedure to derive masses and radii for our two host stars. In each case, we analyse a single co-added spectrum using {\sc SpecMatch-emp} (Yee, Petigura \& von Braun 2017), to determine the stellar effective temperature, \teff, the stellar radius, \rstar, the stellar metallicity, \feh, and the stellar surface gravity, \logg. {\sc SpecMatch-emp} compares a stellar spectrum to spectra from a library of well-characterised stars. This stellar library contains 404 stars ranging from F1 to M5 in spectral type, which have high-resolution ($R \approx 60\,000$) Keck/HIRES spectra, as well as properties derived from other observations (interferometry, asteroseismology, spectrophotometry) and from LTE spectral synthesis.

The uncertainties on the radii from {\sc SpecMatch-emp} are relatively large, particularly in the case of the hotter \celeven. We therefore instead choose to use the \teff and \feh values from {\sc SpecMatch-emp}, and the stellar density, \densstar, determined from the transit light curves (Section~5), as inputs to the empirical relations of Southworth (2011). These relations are based on 90 detached eclipsing binary systems, and can be used to compute the stellar mass and radius. The masses and radii derived in this way are reported, along with the  temperatures and metallicities from {\sc SpecMatch-emp}, in Table~3. We use {\it Gaia} parallaxes to derive stellar radii as a check of the above method, but do not adopt these values, since the {\it Gaia} extinction values are unreliable at the individual-star level (Andrae et al. 2018). 

\subsection{\ceight}
\label{sec:spec_c8}

For \ceight, we used a co-added spectrum comprised of the seven FIES spectra. The stellar radius value from the {\sc SpecMatch-emp} analysis is $0.72 \pm 0.07$~\rsol, which is in excellent agreement with the value derived using Southworth's empirical relations (Section~3.1, Table~3). 

A spectral analysis was also performed on the Tull reconnaissance  spectrum, using {\sc Kea} (Endl \& Cochran 2016), yielding the following parameters: $\teff = 4680 \pm 97$~K, $\logg = 4.38 \pm 0.16$~(cgs), $\feh = -0.24 \pm 0.10$, and $\vsini = 2.2 \pm 0.3$~\kms. The \teff and \logg values are in reasonable agreement (within 2$\sigma$) with those from {\sc SpecMatch-emp}, although we note that the metallicity values differ by more than 2$\sigma$.

Petigura et al. (2018) report stellar parameters for \ceight, based on a Keck/HIRES spectrum. We find that our values are in excellent agreement with theirs (\teff = $4398 \pm 70$~ K, \feh = $0.17 \pm 0.12$, and \rstar = $0.73 \pm 0.1$~\rsol). Finally, we used the parallax value from the second data release of the Gaia mission (Gaia Collaboration et al. 2016; 2018), corrected with the systematic offset derived by Stassun \& Torres (2018), along with the bolometric correction ($BC_G = -0.236 \pm 0.013$ mag) of Andrae et al. (2018) and our \teff value to estimate the stellar radius of \ceight, assuming zero extinction. We derive a radius of $0.74 \pm 0.03$~\rsol, which is in good agreement with our adopted value.

\subsection{\celeven}
\label{sec:spec_c11}

The seven HARPS spectra of \celeven were co-added, and analysed using the method described above. The radius derived using {\sc SpecMatch-emp} is $1.36 \pm 0.22$~\rsol, which agrees well with our adopted value (Table~3). 
As a check, we also analysed the same co-added spectrum using {\sc SME} (Spectroscopy Made Easy; Valenti \& Piskunov 1996; Valenti \& Fischer 2005) with {\sc ATLAS 12} model spectra (Kurucz 2013) and pre-calculated atomic parameters from the VALD3 database (Ryabchikova et al. 2011; 2015). The microturbulent velocity was fixed to 1.3~km~s$^{-1}$ (Bruntt et al. 2010), and the macroturbulent velocity to 5.2~km~s$^{-1}$  (Doyle et al. 2014). The results of our {\sc SME} analysis ($\teff = 6220 \pm 120$~K, $\feh = 0.15 \pm 0.15$, \logg = $4.28 \pm 0.12$) are also in excellent agreement with our adopted values. 

A further comparison was made to the stellar parameters available at the `ExoFOP-K2' website\footnote{https://exofop.ipac.caltech.edu/k2/edit\_target.php?id=229426032} which were generated using the methodology of Huber et al. (2016). These parameters have very much larger uncertainties than our parameters, but all parameters except stellar density ($234 \pm 267$~kg m$^{-3}$) agree to within $1 \sigma$. We note that the mass and radius given on ExoFOP result in a higher density of around 420~kg m$^{-3}$. Using the Gaia DR2 parallax, assuming zero extinction, and $BC_G = 0.076 \pm 0.034$ mag, we find \rstar = $1.21 \pm 0.06$~\rsol, which is slightly more than $2\sigma$ from our adopted value. In order to make the Gaia-derived radius match our adopted radius, we require extinction in the Gaia bandpass, $A_G = 0.29$, which is consistent with the value of $0.25^{+0.15}_{-0.20}$ reported in Gaia DR2.

We also compared our stellar parameters to those derived by Soto et al. (2018). The mass and radius estimates are in reasonably good agreement, with the values of  Soto et al. (2018) around $1\sigma$ larger than ours. This is probably explained by the higher temperature found by Soto et al. (2018) (\teff = $6257 \pm 100$~K); using this temperature and our stellar density as inputs to the Southworth (2011) relations, we get a stellar radius very close to their value ($1.42 \pm 0.04$~\rsol cf. Soto et al.'s $1.43^{+0.06}_{-0.07}$~\rsol). We note, however, that the stellar density implied by the Soto et al. (2018) mass and radius values (\densstar = $0.44 \pm 0.06$) is inconsistent at more than $5\sigma$ with their quoted density value (\densstar = $0.102^{+0.012}_{-0.010}$. In solar units, our derived stellar density (from light curve modelling) is \densstar = $0.47\pm0.03$~\denssol. It is unclear how the density quoted by Soto et al. (2018) was derived.


We computed a Lomb-Scargle periodogram using the light curve further decorrelated using a polynomial fit, and with in-transit points removed. We found a peak at around 5.1~d, which we attribute to stellar rotation. The amplitude of this rotational variability varies over the course of the \ktwo observations, and was strongest in the first part of the light curve.

This detected period closely matches that found by Soto et al. (2018) ($5.07\pm0.02$~d). Using their period and our stellar radius and \vsini (from {\sc SME}) values, we determine the stellar inclination angle, $i_* = 59^{+9}_{-7}$ degrees. This is slightly larger than the $51.56^{+3.73}_{-2.80}$ degrees determined by Soto et al. (2018), and we also note that our $2\sigma$ error bar ($59^{+31}_{-13}$~degrees) encompasses $90^\circ$. We would therefore caution against concluding that the stellar spin and planetary orbital axes are misaligned; our smaller stellar radius is consistent with them being aligned or near-aligned.

\subsection{Distances}

The distances quoted in Table~3 are derived from the parallaxes listed in the second GAIA data release (Gaia Collaboration et al. 2016; 2018). They are in good agreement with distances calculated from estimates of the absolute magnitude, albeit with significantly smaller uncertainties. In particular, we note that the Gaia distance to \celeven is consistent with that derived by Soto et al. (2018), but that the Gaia uncertainty is approximately 20 times smaller.

\subsection{Ages}

\begin{figure}
    \centering
	\includegraphics[width=6.3cm]{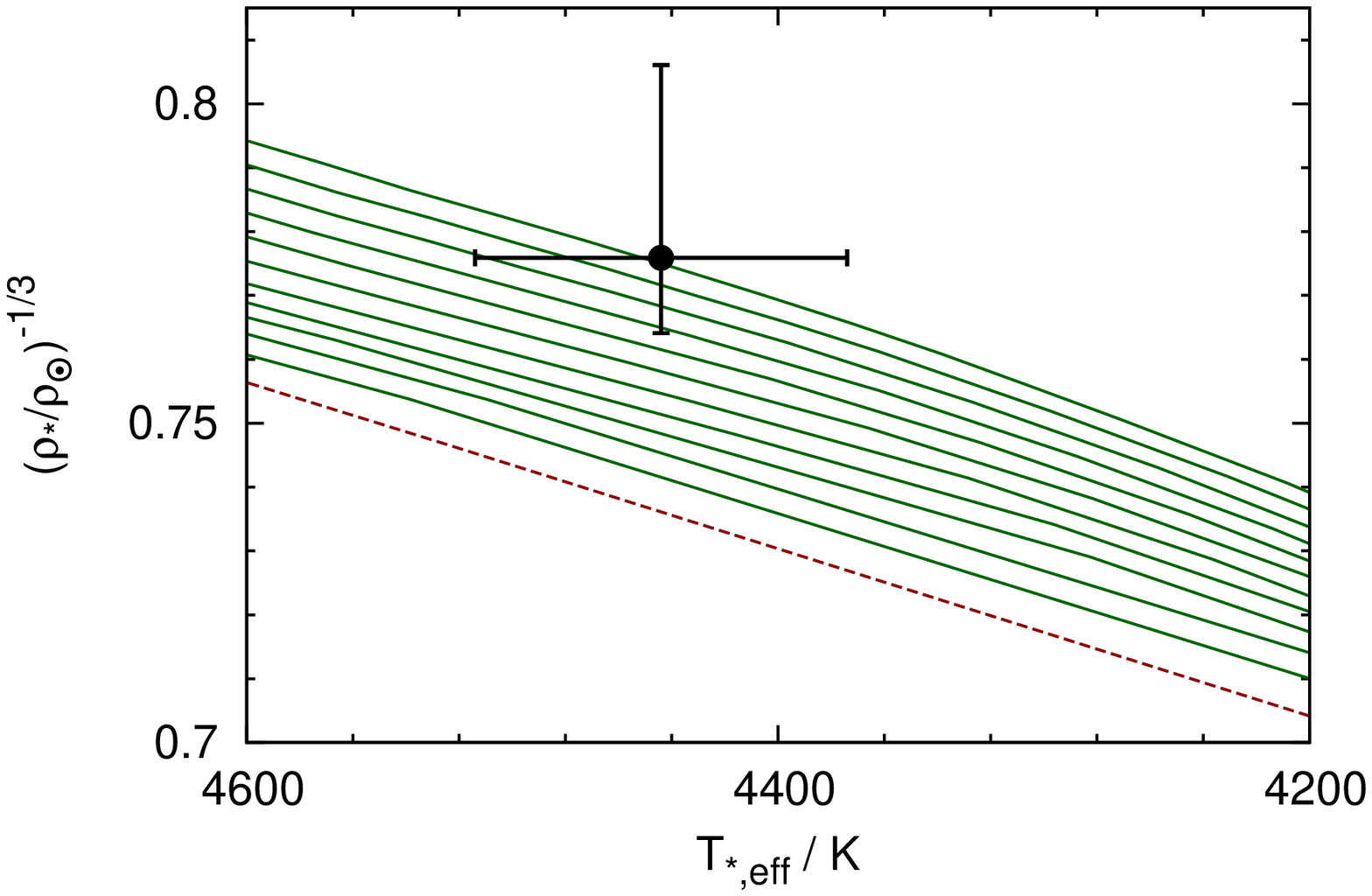}  \includegraphics[width=6.3cm]{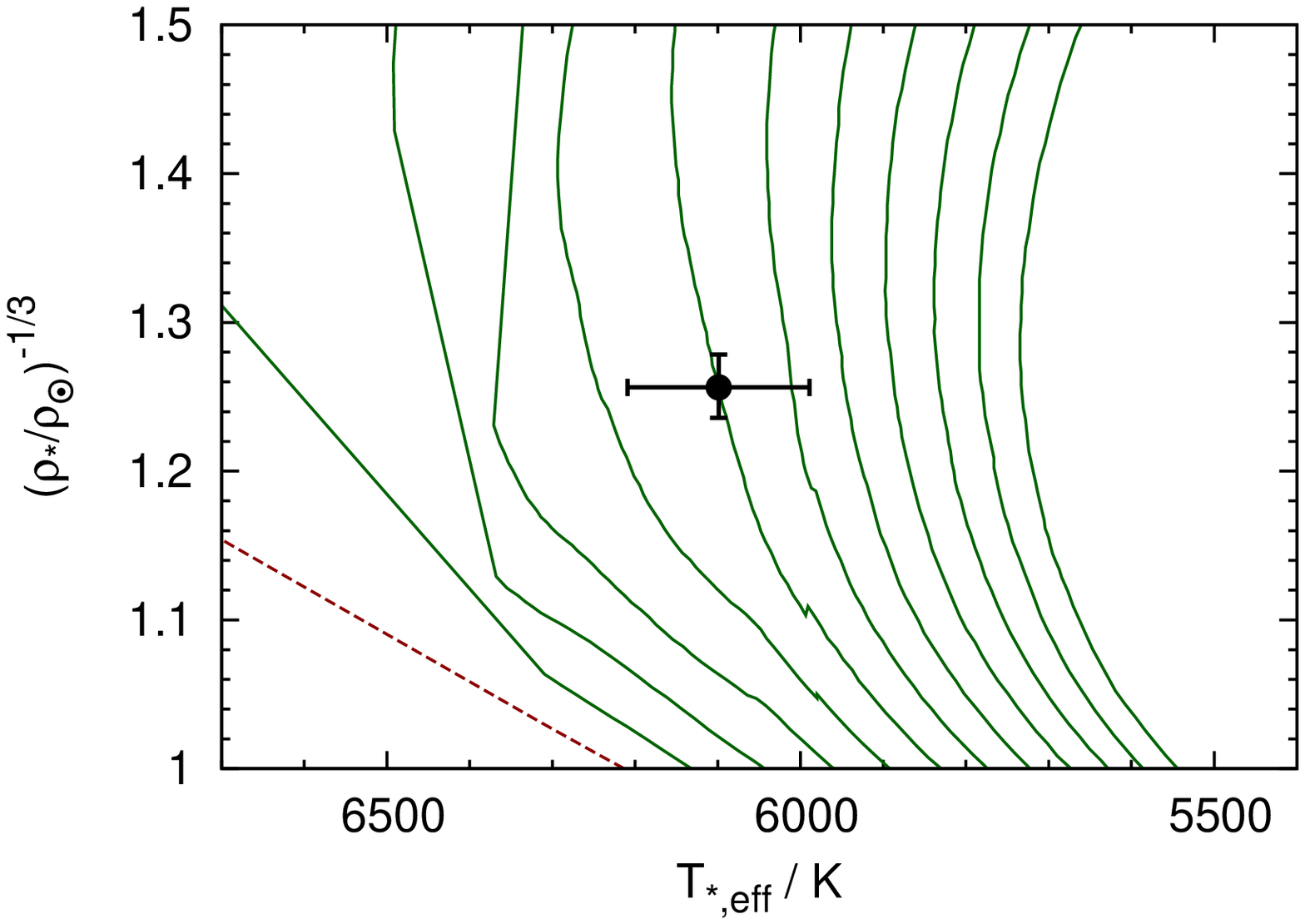}
    \caption{Modified Hertzsprung-Russell diagrams for \ceight (left) and \celeven (right). In each panel the target star is represented with a black circle. Dartmouth isochrones (Dotter et al. 2008) are shown for 1.0~Gyr (dashed red line), and at 1~Gyr spacings, up to 12.0~Gyr.}
    \label{fig:c8_hrd}
\end{figure}

We plotted \ceight alongside Dartmouth isochrones (Dotter et al. 2008) interpolated to our stellar metallicity value (Table~3) at intervals of 1~Gyr (Fig.~1, \textit{left}). We determined the range of ages which are compatible with our stellar density and effective temperature from a simple visual inspection (as was used in e.g. Smith et al. 2014). The age is poorly-constrained - the $1\sigma$ uncertainties span all ages greater than about 8.5~Gyr. We note, however, that these uncertainties are probably underestimated, since we do not account for the uncertainty on the metallicity, nor the systematic errors in the Dartmouth stellar models. The latter effect could be militated against by considering a variety of stellar models (as in e.g. Southworth 2009), but considering the large uncertainties opt not to do this, and instead draw no firm conclusions about the age of \ceight.


Plotting \celeven alongside theoretical isochrones (Dotter et al. 2008) yields a best-fitting age of approximately $6\pm1$~Gyr (Fig.~1, \textit{right}).  As for \ceight, we acknowledge that the uncertainty on this age determination is probably underestimated. Using the $5.07 \pm 0.02$~d rotation period (Soto et al. 2018) as an input to the gyrochronology relation of Barnes (2010), we derive an age for \celeven of $1.2\pm0.7$~Gyr. We use our stellar mass value and linear interpolation of Table~1 of Barnes \& Kim (2010) to determine the convection turn-over timescale. We note, however, that ages derived from isochrones and from gyrochronology often disagree for planet-host stars (Brown 2014; Maxted, Serenelli \& Southworth 2015), perhaps because hot Jupiters tidally interact with their host stars, spinning them up.


\subsection{Spectral type}

The spectral types listed in Table~3 were determined using the tabulation of Pecaut \& Mamajek (2013). Our stellar effective temperatures were compared to those listed in the online version of their table\footnote{http://www.pas.rochester.edu/\~emamajek/EEM\_dwarf\_UBVIJHK\_colors\_Teff.txt}.

\section{Contamination from neighbouring objects}
\subsection{\ceight}

The Subaru/IRCS data were reduced following the procedure in Hirano et al. (2016), and we obtained the calibrated combined images for the saturated and unsaturated frames respectively. To estimate the achieved contrast of the saturated image, we computed the flux scatter within the annulus as a function of angular separation from the centroid of the star. Fig.~2 plots the $5\sigma$ contrast curve together with the target image with the field-of-view of $4^{\prime\prime}\times4^{\prime\prime}$. \ceight is a single star to the detection limit, meaning that the light curve is free from contamination from nearby objects.

\begin{figure}
    \centering
	\includegraphics[width=8cm]{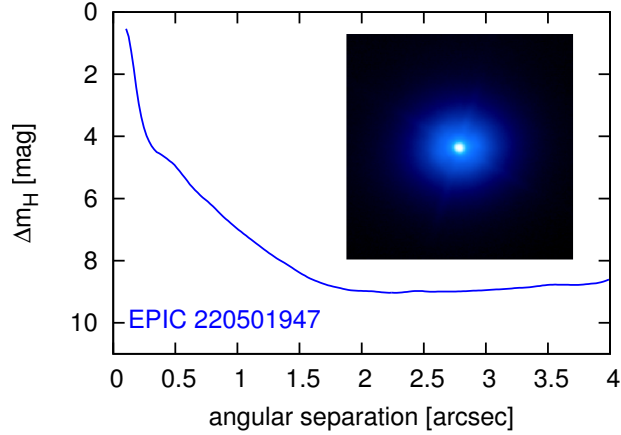}
    \caption{Results of Subaru IRCS imaging of \ceight. The curve indicates the $5\sigma$ detection limit, as a function of angular separation, and the inset image (4$^{\prime\prime} \times 4^{\prime\prime}$, North is up, East is left) indicates that there is no evidence for any close companions to \ceight.}
    \label{fig:c8_img}
\end{figure}

\subsection{\celeven}

The MuSCAT imaging reveals \celeven to be in a rather crowded field, with several faint objects nearby. Using the $r\prime$ band image (Fig.~3), we detected a total of ten objects fainter than the target within the photometric aperture used to generate the light curve. The total flux contribution of these objects relative to the target flux is 0.042. We adopt this value for the quantity of contaminating `third' light, and conservatively estimate an uncertainty of half, i.e. 0.021 -- to account for measurement errors and the difference between the Kepler and $r\prime$ bandpasses. {The third light is accounted for in our modelling of the transit light curve (Section~\ref{sec:tlcm})}, and has the effect of changing the planet radius at approximately the $1\sigma$ level.

\begin{figure}
	\centering
	\includegraphics[width=6cm]{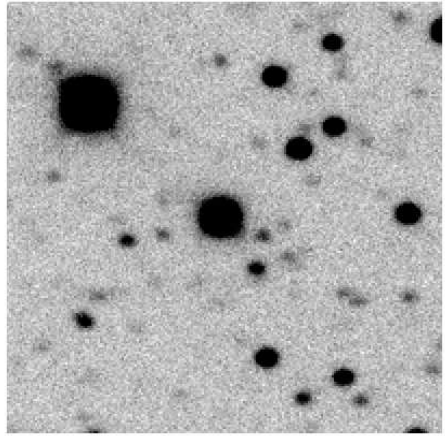}
    \caption{MuSCAT r-band image, centred on \celeven. North is up, East is to the left, and the image is $72^{\prime \prime} \times 72^{\prime \prime}$. A number of faint contaminating stars can be seen in the close vicinity of the target.}
    \label{fig:c11_img}
\end{figure}

We constructed a high-resolution image by co-adding the best thirty per cent of the  TCS/FastCam images, giving a total exposure time of 150~s. The typical Strehl ratio of these images is about 0.07. In order to construct the co-added image, each individual frame was bias-subtracted, aligned and co-added and then processed with the FastCam dedicated software developed at the Universidad Polit\'ecnica de Cartagena (Labadie et al. 2010; J\'odar et al. 2013). Fig.~4 shows the contrast curve that was computed based on the scatter within the annulus as a function of angular separation from the target centroid. 

Three neighbouring objects were found in the image, at separations from \celeven of between 7$^{\prime\prime}$ and 11$^{\prime\prime}$. The relative fluxes of these objects are consistent with those determined by MuSCAT. No bright companions were detected within 7$^{\prime\prime}$ of the target (Fig.~4).

\begin{figure}
    \centering
    \includegraphics[width=8cm]{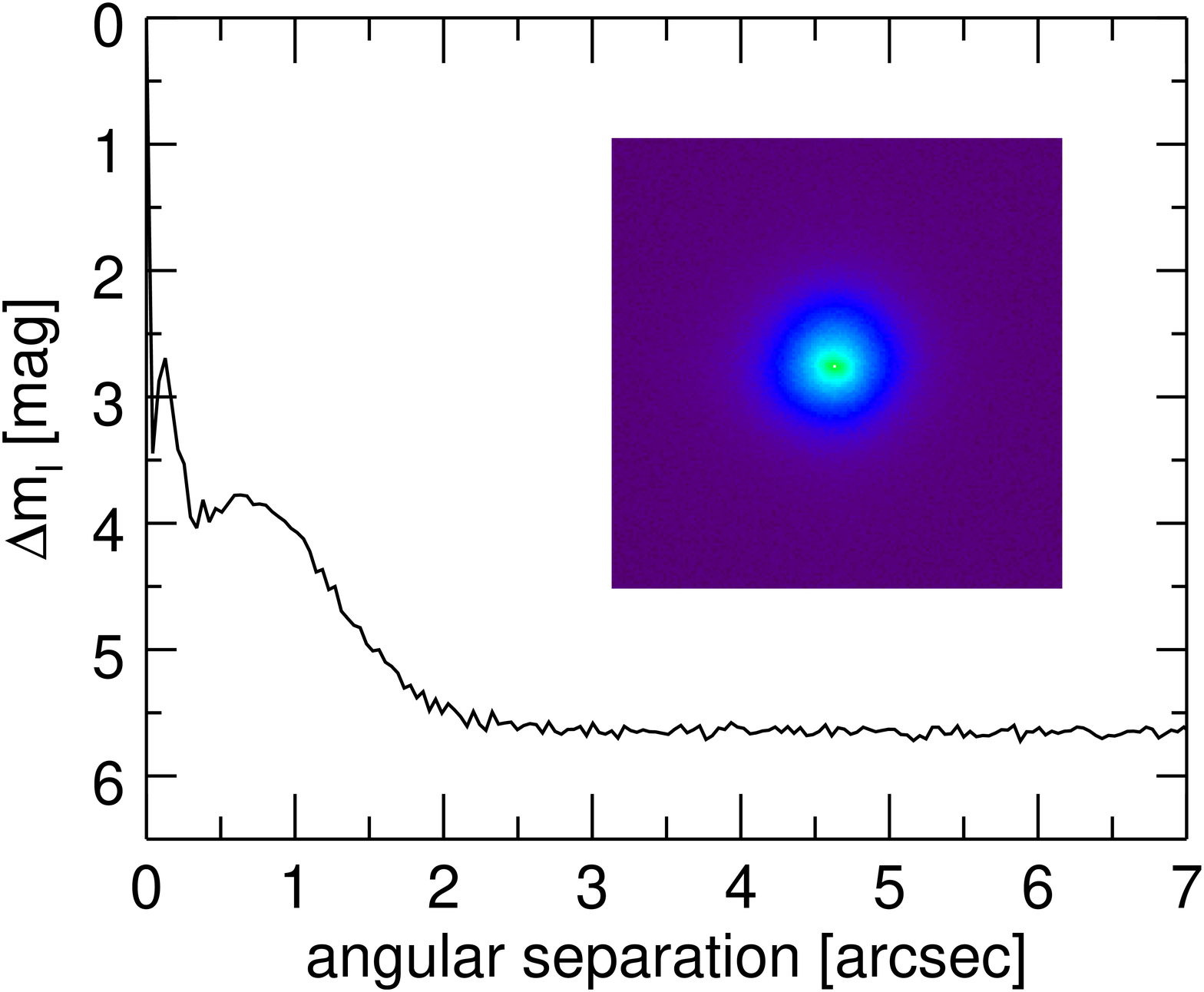}
    \caption{I-band magnitude contrast curve as a function of angular separation up to $7.0"$ from \celeven obtained with the FastCam camera at TCS. The solid line indicates the $5\sigma$ detection limit for the primary star. The inset shows the $7"\times 7"$ combined image of \celeven. North is up and East is left.}
    \label{fig:lucky}
\end{figure}

\section{Determination of system parameters}
\label{sec:tlcm}

\begin{figure}
    \centering
	\includegraphics[width=12cm]{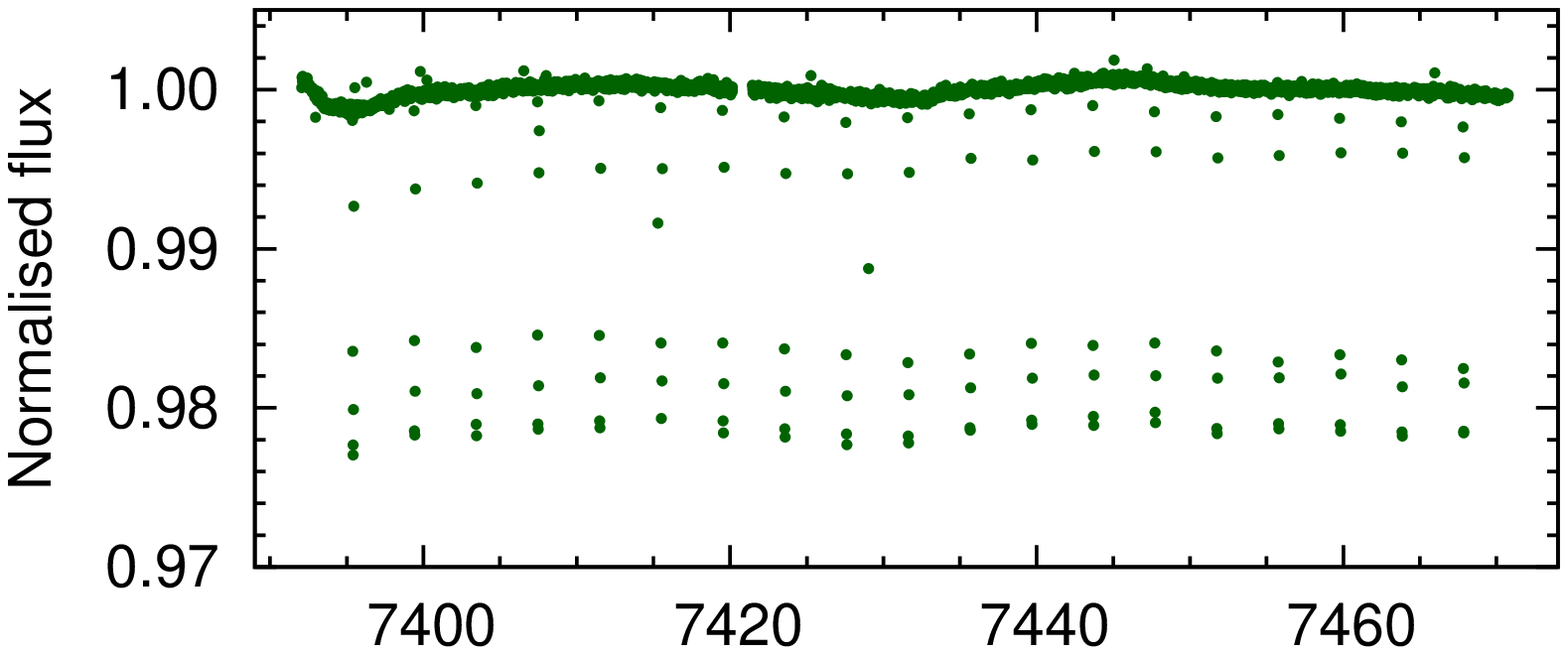}
	\includegraphics[width=12cm]{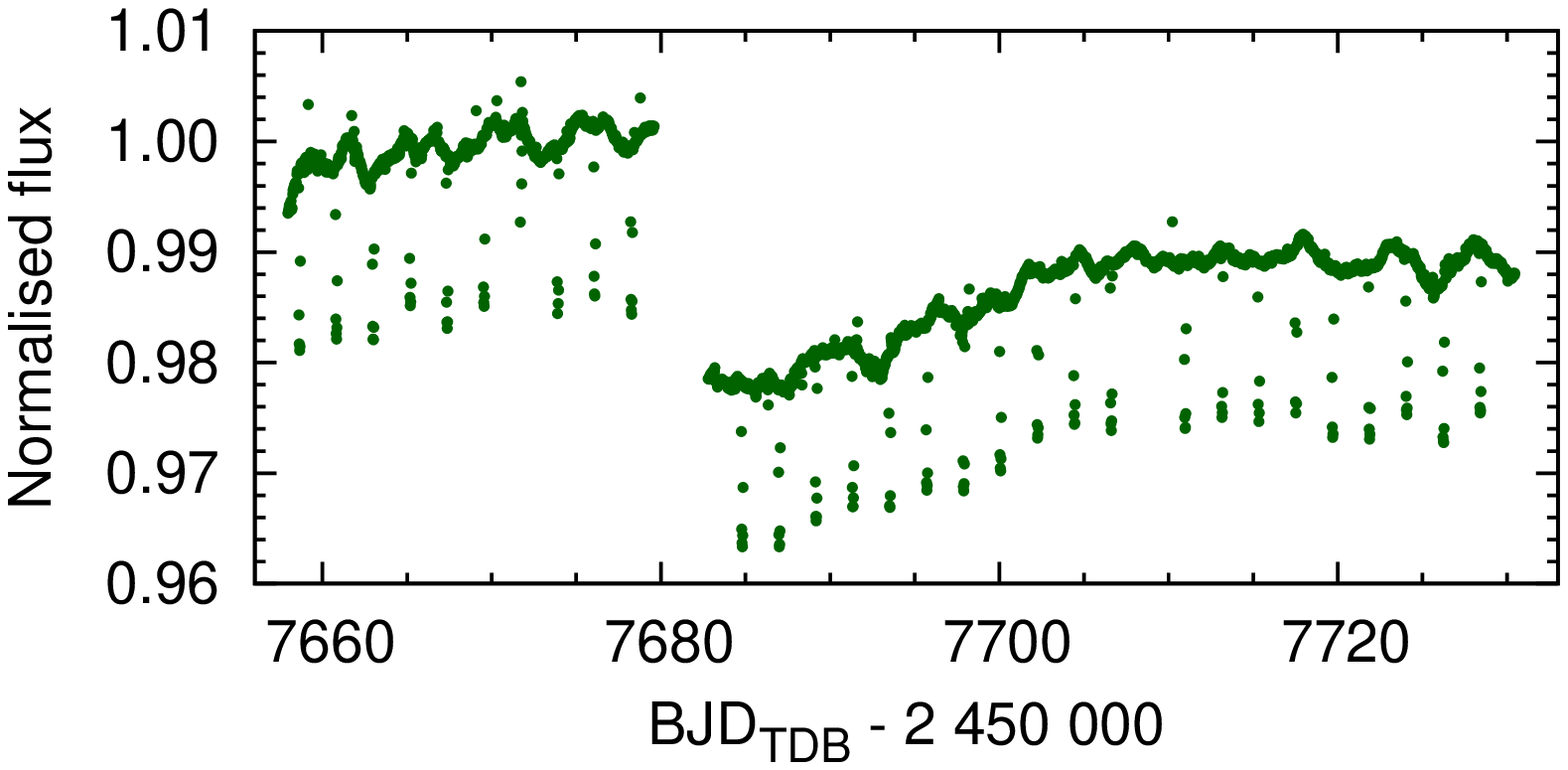}
    \caption{\ktwo light curves for \ceight (upper panel) and \celeven (lower panel). The \ceight light curve was produced using the {\sc Everest} code  (Luger et al. 2016), and the \celeven light curve by Andrew Vanderburg (following Vanderburg \& Johnson 2014). The discontinuity in the lower panel is a result of a change in the roll angle of \ktwo during Campaign 11 (see Section~2.1 for further details).}
    \label{fig:raw_lcs}
\end{figure}

\begin{figure}
    \centering
	\includegraphics[width=8cm]{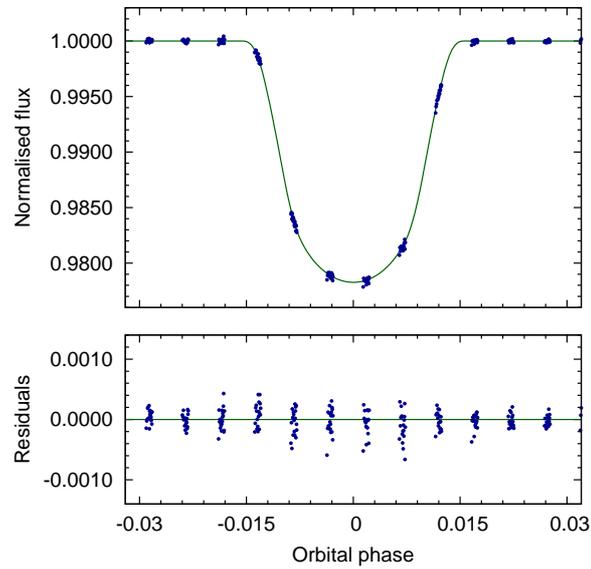}
    \caption{Phase-folded \ktwo photometry (blue circles) and best-fitting model (solid green line) for \ceight, with residuals to the model shown in the lower panel. The light curve is that of Luger et al. (2016).}
    \label{fig:c8_lc}
\end{figure}

\begin{figure}
    \centering
	\includegraphics[width=8cm]{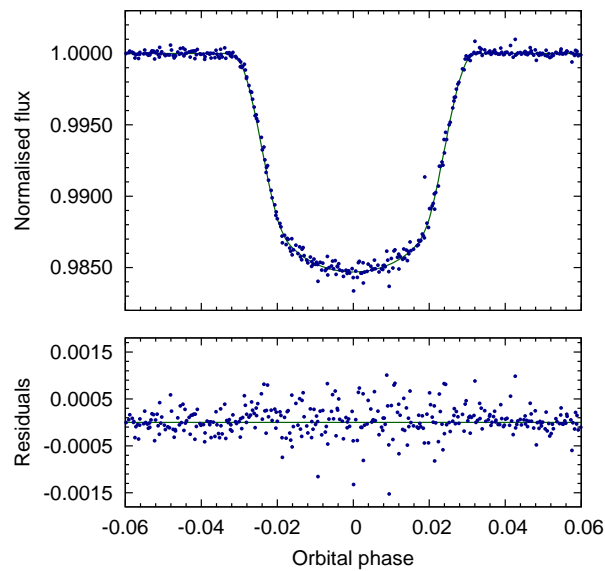}
    \caption{Phase-folded \ktwo photometry (blue circles) and best-fitting model (solid green line) for \celeven, with residuals to the model shown in the lower panel. The light curve is that of Vanderburg \& Johnson (2014).}
    \label{fig:c11_lc}
\end{figure}

\begin{figure}
    \centering
	\includegraphics[width=8cm]{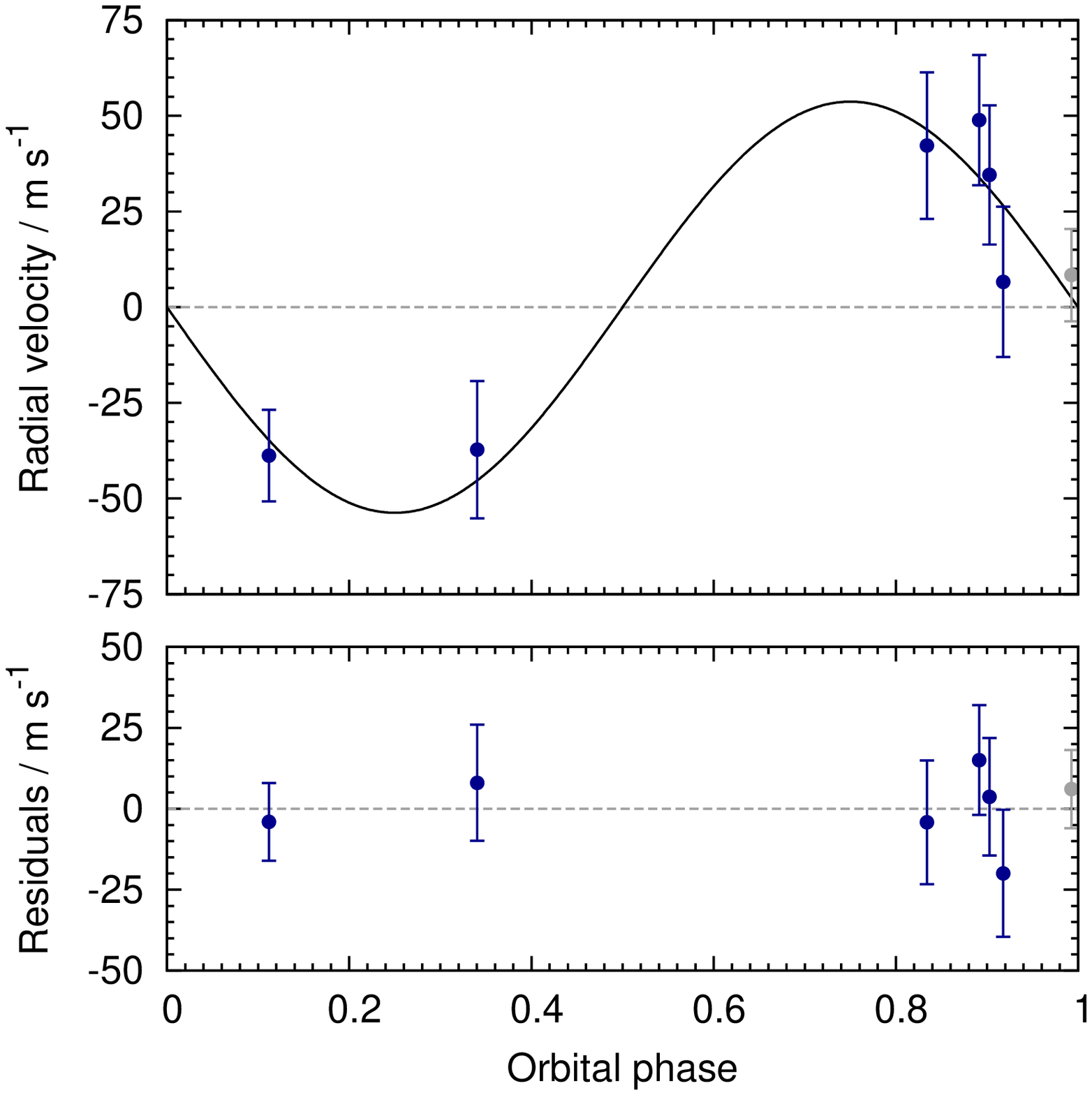}
    \caption{Radial velocities from FIES for \ceight. The RV point taken during transit is shown in grey. Our best-fitting model is shown with a solid black line, and the residuals to the model are plotted in the lower panel. The data are phase-folded, and the systemic radial velocity, $\gamma$, has been subtracted.}
    \label{fig:c8_rv}
\end{figure}

\begin{figure}
    \centering
	\includegraphics[width=8cm]{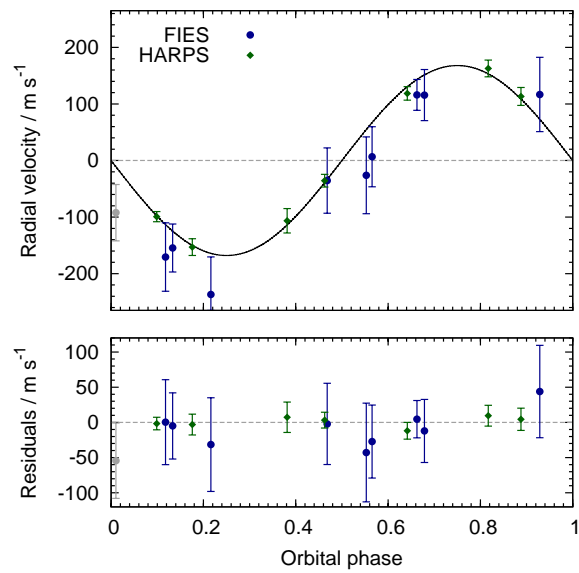}
    \caption{Radial velocities from FIES (blue circles) and HARPS (green squares) for \celeven. The FIES RV point taken during transit is shown in grey. Our best-fitting model is shown with a solid black line, and the residuals to the model are plotted in the lower panel. The data are phase-folded, and the systemic radial velocity, $\gamma$, has been subtracted.}
    \label{fig:c11_rv}
\end{figure}

\begin{figure}
    \centering
	\includegraphics[width=8cm]{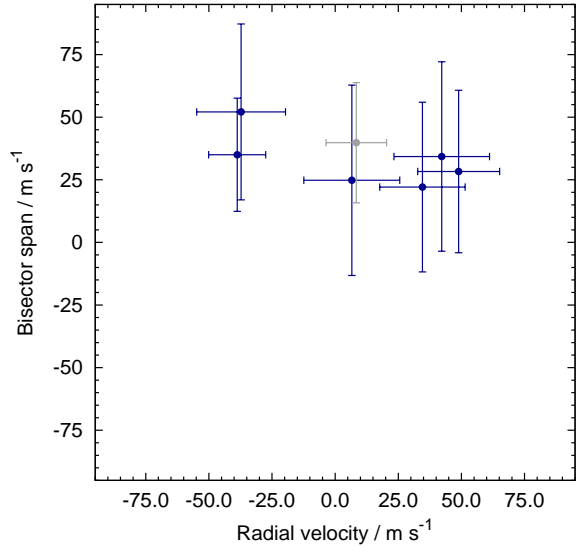}
    \caption{Bisector span as a function of radial velocity for \ceight. As in Fig.~8, the systemic radial velocity, $\gamma$, has been subtracted. The uncertainties in the bisector spans are taken to be twice those of the radial velocities. The RV point taken during transit is shown in grey.}
    \label{fig:c8_bis}
\end{figure}

\begin{figure}
    \centering
	\includegraphics[width=8cm]{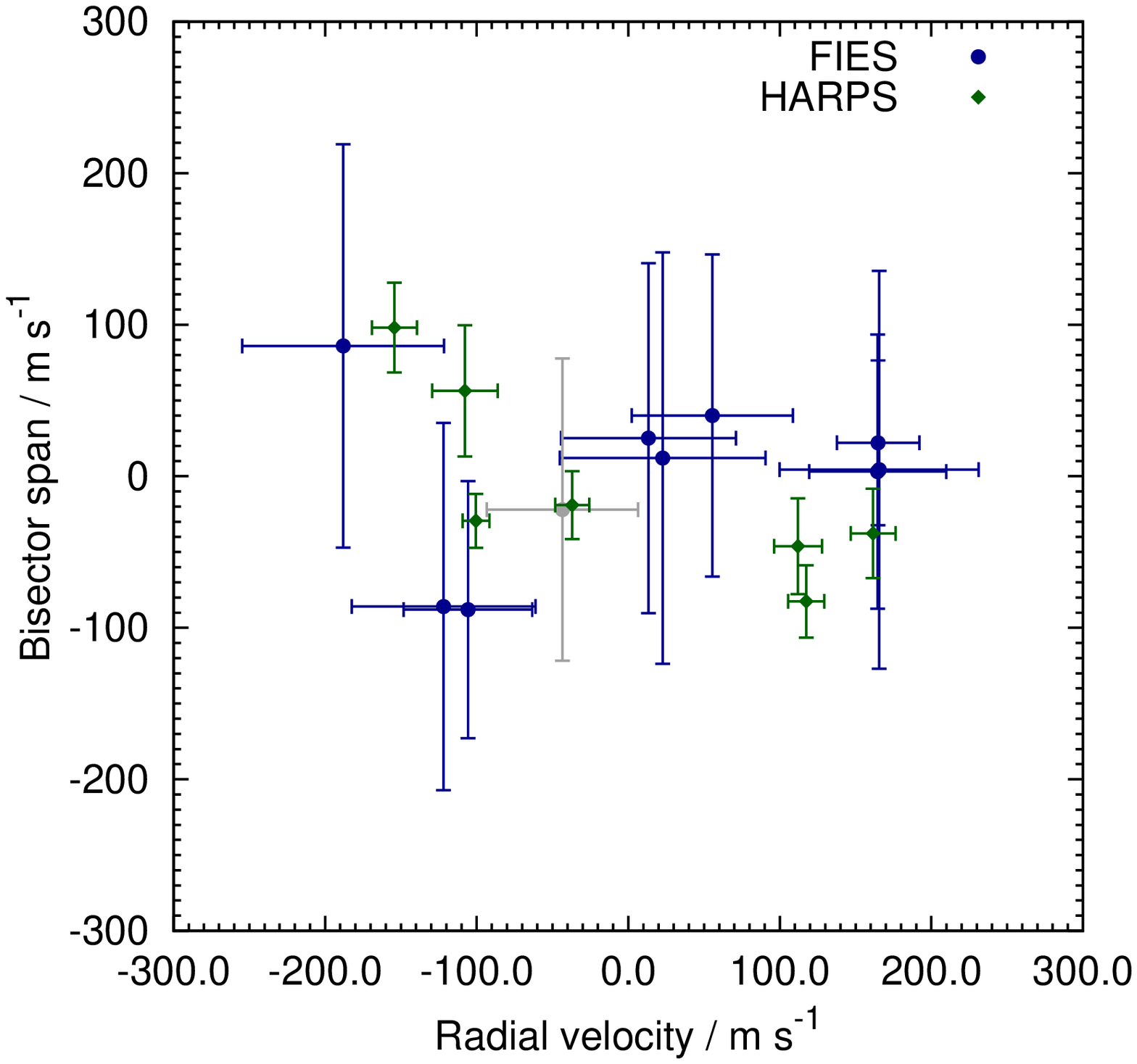}
    \caption{Bisector span as a function of radial velocity for \celeven. As in Fig.~9, the systemic radial velocity, $\gamma$, has been subtracted, as has the fitted RV offset between the FIES and HARPS instruments, $\gamma_{\mathrm{F-H}}$. The uncertainties in the bisector spans are taken to be twice those of the radial velocities.}
    \label{fig:c11_bis}
\end{figure}

\subsection{Light curve preparation}

We use the {\sc Everest} (Luger et al. 2016) \ktwo light curve for \ceight (Fig.~5, upper panel)\footnote{We have previously found that the {\sc Everest} and Vanderburg \& Johnson (2014) light curves are of very similar quality, with the {\sc Everest} curves containing very slightly less noise on average. We therefore use the {\sc Everest} light curve for \ceight. For \celeven, however, there was no {\sc Everest} curve available when we started our modelling efforts, so we use the Vanderburg \& Johnson (2014) curve instead.}. For modelling the transit, we cut the light curve into pieces, selecting only those light curve points within two transit durations of the transit midtime for modelling. This makes detrending the light curve for stellar activity more straightforward, as well as reducing model computation times. This results in a series of light curve sections of length $4\, T_{\rm 14}$ (approximately 10 hours in the case of \ceight), centred on the midpoint of each transit. Each section of the light curve is detrended using a quadratic function of time to remove the remaining signatures of stellar variability. Finally, we remove three obvious outliers from the light curve.


For \celeven, we perform the same procedure as above, but we instead use the light curve of Vanderburg \& Johnson (2014) (lower panel of Fig.~5). In addition to the transits, the light curve exhibits a quasi-periodic signal which we attribute to stellar rotational variability and investigate further in Section~3.3.

\subsection{The TLCM code}

We model each system using the {\sc Transit and Light Curve Modeller (TLCM)} code. {\sc TLCM} has been used to model exoplanet light curves and radial velocities in numerous previous studies, including planets discovered in long-cadence \ktwo data (e.g. K2-99b, Smith et al. 2017). The code is described in Csizmadia et al. (2015), and a more detailed description will accompany the first public release of the code (Csizmadia, under review).

In brief, TLCM fits the photometric transit using the Mandel \& Agol (2002) model, compensating for \ktwo's long exposure times using numerical integration, and simultaneously fits a Keplerian orbit to the RV data. {\sc TLCM} uses the combination of a genetic algorithm to find the approximate global minimum, followed by simulated annealing and Markov-chain Monte Carlo phases to refine the solution, and explore the neighbouring parameter space for the determination of uncertainties on the model parameters.

\subsection{Combined fit}
\label{sec:basic_fit}

For our basic fit, we fit for the following parameters: the orbital period, $P$, the epoch of mid-transit, $T_0$, the scaled semi-major axis ($a/\rstar$), planet-to-stellar radius ratio ($\rplanet / \rstar$), the impact parameter, $b$, the limb-darkening parameters, $u_+$ and $u_-$ (see below), the systemic stellar RV, $\gamma$, and the RV semi-amplitude, $K$. In the case of \celeven, for which we have RV data from FIES and HARPS, we also fit the systematic offset between these two instruments, $\gamma_\mathrm{F-H}$. For each system, one of our RV measurements was taken during transit. Since we do not model the Rossiter-Mclaughlin effect, these points are not included in the modelling, and are marked in italics in Table~2.

\subsection{Limb darkening}
\label{sec:ld}
Limb-darkening is parametrized using a quadratic model, whose coefficients, $u_a$ and $u_b$ are transformed to the fit parameters $u_+ = u_a + u_b$ and $u_- = u_a - u_b$. In the case of \celeven, $u_+$ and $u_-$ are free parameters. For \ceight, the observational cadence of \ktwo is close to an integer fraction of the orbital period. This results in clumps of data points in phase space, rather than the data being evenly distributed in phase (Fig.~6). Transit ingress and egress are poorly covered, providing a weaker constraint on the limb-darkening parameters than would otherwise be the case. We therefore opt to constrain the limb-darkening parameters to take values close to ($\pm 0.01$) the theoretical values of Sing (2010) for the relevant stellar parameters and the {\it Kepler} bandpass ($u_+ = 0.7349$, $u_- = 0.5689$). We discuss this issue and related problems arising from the poor coverage of ingress and egress in an Appendix to this paper.

\subsection{Orbital eccentricity}
\label{sec:ecc}
In our basic fit, we fix the orbital eccentricity to zero, but we also used {\sc TLCM} to fit for the orbital eccentricity, $e$, rather than forcing a circular orbital solution. The additional parameters we fit for in this case are \ecos and \esin, where $\omega$ is the argument of periastron. We used the $\chi^2$ values of the resulting fits to calculate the Bayesian Information Criterion (BIC) in order to establish whether the improved RV fit justifies the additional model parameters.

For both systems, we found a larger BIC value for the eccentric fit (for \ceight $\mathrm{BIC}_\mathrm{ecc} - \mathrm{BIC}_{e = 0} = 3.6$, and for \celeven, $\mathrm{BIC}_\mathrm{ecc} - \mathrm{BIC}_{e = 0} = 5.0$). For the purposes of calculating the BIC, we considered the number of data points to be the number of RV points only, since these provide most of the information regarding orbital eccentricity. Including the photometric data points in the total would increase the BIC values, making a circular orbit even more favourable.  We note that for  neither system is the best-fitting eccentricity found to be significant at the $3\sigma$ level, although the eccentricity of \ceight is poorly-constrained because of the incomplete phase coverage of the RV data. We therefore adopt $e = 0$ for both systems, as expected given both theoretical predictions for close-in exoplanetary systems, and empirical evidence that such planets only rarely exist on significantly eccentric orbits (e.g. Anderson et al. 2012).

\subsection{RV drift}

We also tried fitting for a linear trend in the radial velocities of each star, the presence of which can be indicative of the presence of a third body in the system. In both cases, we found that the best-fitting radial acceleration is not significant, and that the BIC clearly favours the simpler model. In summary, there is no evidence for the presence of a third body in either system.

\subsection{Checks for a blended binary system}

A blended eclipsing binary can mimic a transiting planetary system, but will exhibit a correlation between the RV and the RV bisector spans (Queloz et al. 2001). In Figs.~10 and 11, we plot these two quantities, and find that there is no such correlation in either case, as expected for true planetary systems. We note, however, that G\"unther et al. (2018) determine that the lack of a bisector correlation does not rule out all blend scenarios. 

We also look for a variation of transit depth with photometric aperture size -- a powerful discriminant which has been used previously to disprove previously-validated planetary candidates from \ktwo (Cabrera et al. 2017). For \ceight, we observe no change in transit depth with increasing aperture radius. For \celeven, we observe the transit depth shrink with increasing aperture radius; this is the expected behaviour given that there is a nearby companion whose light dilutes the transit depth when a large photometric aperture is used. We further note that our high-resolution imaging rules out the presence of a binary companion capable of mimicking the observed transit signals, unless such a companion lies at very small ($\lesssim 1^{\prime\prime}$) sky-projected separations. Estimating the probability of such a scenario is non-trivial, requiring detailed simulations which are beyond the scope of this paper.

\subsection{Additional photometric signals}

We tried fitting for an occultation (centred on phase 0.5, given the evidence for circular orbits in both systems). No evidence was found for the presence of an occultation signal in the light curve of either system. Similarly, we found no compelling evidence of any transit timing variations (TTV) in either system. Interestingly, there seems to be a variation in the transit depth of \celeven~b. This explains the higher in-transit residual scatter observed in Fig.~7, which we suggest is caused by stellar spots, which are also responsible for the rotational modulation seen in the light curve (Section~3.3).

\subsection{Additional RV data}
~\\
The RV semi-amplitude and planet mass that we determine for \celeven differ somewhat from the values ($K = 210 \pm 10$~ms$^{-1}$; \mplanet$ = 1.60 \pm 0.11$~\mjup) reported by Soto et al. (2018). We tried including their RVs (four measurements from HARPS, and nine from CORALIE) in our fit, and found that we require an offset between our HARPS measurements and theirs. We suggest that the need for this arises from the different reduction pipelines used to obtain the RVs from the HARPS spectra. Including the RVs of Soto et al. (2018) yields $K = 180^{+5}_{-8}$~ms$^{-1}$, which is compatible (at the $\approx 1\sigma$ level) with the value obtained from our data alone ($K = 168^{+5}_{-3}$~ms$^{-1}$; Table~4), but almost $3\sigma$ from the value of Soto et al. (2018). The source of this apparent discrepancy is unclear.

\section{Conclusions}

In summary, we find that the planet orbiting the K5 dwarf \ceight in an approximately 4-d orbit is slightly larger and more massive than Saturn ($1.06 \pm 0.01$~\rsat and $1.12 \pm 0.21$~\msat). The planetary parameters (transit duration, impact parameter, and planetary radius) reported in the planet candidate list of Petigura et al. (2018) are in good agreement with those derived in our analysis. The radius of \ceight~b seems to be fairly typical for a hot Saturn, slightly smaller than the similar HATS-6~b and WASP-83~b (Hartman et al. 2015; Hellier et al. 2015), but significantly larger than that of the anomalously dense HD~149026~b (Sato et al. 2005), which is thought to be extremely metal-rich (Speigel, Fortney \& Sotin 2014).

We confirm the conclusion of Soto et al. (2018), that \celeven~b is significantly inflated. We find that the planet is typical of an inflated hot Jupiter -- slightly more massive than Jupiter, but with a radius some 60 to 70 per cent larger than the largest planet in the Solar System. The planet orbits an F8 dwarf star.

\begin{sidewaystable}
\caption{System parameters from {\tt TLCM} modelling} 
\label{tab:tlcm}
\begin{tabular}{lcccc}
\hline
\hline
Parameter & Symbol & Unit & \ceight & \celeven\\
\hline 
\textit{{\tt TLCM} fitted parameters:} &&\\
&\\
Orbital period	    	    	    	    & $P$ & d & $4.024867 \pm 0.000015$ & $ 2.1805577 \pm 0.0000057$ \\
Epoch of mid-transit	    	    	    & $T_{\rm 0}$ &$\mathrm{BJD_{TDB}}$ & $ 2457395.4140498 \pm 0.0000012 $ & $ 2457656.4633789\pm0.0000048$ \\
Scaled orbital major semi-axis              & $a/R_{\rm *}$ &--& {}$ 13.76^{+0.19}_{-0.44}$ & $ 5.503^{+0.015}_{-0.207} $ \\
Ratio of planetary to stellar radii         & \rplanet / \rstar &--& $0.1304^{+0.0014}_{-0.0007}$  & $ 0.1195^{+0.0015}_{-0.0005} $ \\
Transit impact parameter                    & $b$ &--& $0.17^{+0.14}_{-0.11}$ & $ 0.520^{+0.057}_{-0.002}$ \\
Limb-darkening parameters                   & $u_+$ &--&$ 0.734 \pm 0.007^{*}$  & $ 0.603^{+0.068}_{-0.062}$ \\
						                    & $u_-$ &--&$ 0.569 \pm 0.007^{*}$ & $ 0.02^{+0.11}_{-0.20}$ \\
Stellar orbital velocity semi-amplitude 	& $K$ &m s$^{-1}$ & $54 \pm 10 $ & $167.9^{+4.7}_{-3.1}$\\
Systemic radial velocity          			& $\gamma$ &km s$^{-1}$ & $-16.6185 \pm 0.0067$ & $ -22.4700^{+0.0004}_{-0.0147}$\\
Velocity offset between FIES and HARPS 		& $\gamma_{\rm F-H}$ & m s$^{-1}$ & $ -- $ & $ 143^{+15}_{-9}$\\
\\
\textit{Derived parameters:} & \\
& &  \\
Orbital eccentricity (adopted) 	    	    & 	$e$ &...& $0$   & $  0  $ \\
Stellar density     	    	    	    & 	\densstar & kg m$^{-3}$ & $ 3043^{+128}_{-280}$ & $ 663 \pm 40 $ \\
Planet mass 	    	    	    	    & 	\mplanet &\mjup & $0.335 \pm 0.062$  & $ 1.236 \pm 0.044 $ \\
Planet radius	    	    	    	    & 	\rplanet &\rjup & $0.897^{+0.011}_{-0.005} $ & $ 1.642 \pm 0.050 $ \\
Planet density	    	    	    	    & 	\densplanet & kg m$^{-3}$ & $ 612 \pm 115 $ & $ 370 \pm 36 $ \\
Orbital major semi-axis     	    	    & 	$a$ &AU  & $ 0.0451^{+0.0006}_{-0.0014}$ & $ 0.0353 \pm 0.0012 $ \\
Orbital inclination angle   	    	    & 	$i_\mathrm{p}$ &$^\circ$  & $ 89.30^{+0.46}_{-0.62} $ & $  84.6 \pm 0.3 $ \\
Transit duration                            & $T_{\rm 14}$ &d&$0.1041^{+0.0011}_{-0.0007} $ & $ 0.1251 \pm 0.0032 $ \\
Planet equilibrium temperature$^\dagger$ & $ T_{\mathrm{p, eql,}A = 0}$ & K & $852^{+14}_{-6} $ & $  1838 \pm 38 $ \\\hline
\multicolumn{5}{l}{$^*$For \ceight, the limb-darkening coefficients are not freely fitted - see Section~5.4 for details.}\\
\multicolumn{5}{l}{$^\dagger$The equilibrium temperature is calculated assuming a planetary albedo of zero, and isotropic re-radiation.}
\end{tabular} \\ 
\end{sidewaystable}



\Acknow{
This work was supported by the KESPRINT collaboration, an international consortium devoted to the characterisation and research of exoplanets discovered with space-based missions.\\www.kesprint.science This paper includes data collected by the Kepler mission. Funding for the Kepler mission is provided by the NASA Science Mission directorate. Some of the data presented in this paper were obtained from the Mikulski Archive for Space Telescopes (MAST). STScI is operated by the Association of Universities for Research in Astronomy, Inc., under NASA contract NAS5-26555. Support for MAST for non-HST data is provided by the NASA Office of Space Science via grant NNX09AF08G and by other grants and contracts. This research has made use of the Exoplanet Follow-up Observation Program website, which is operated by the California Institute of Technology, under contract with the National Aeronautics and Space Administration under the Exoplanet Exploration Program.

We greatly thank the NOT staff members for their precious support during the observations. Based on observations obtained with the Nordic Optical Telescope (NOT), operated on the island of La Palma jointly by Denmark, Finland, Iceland, Norway, and Sweden, in the Spanish Observatorio del Roque de los Muchachos (ORM) of the Instituto de Astrof\'isica de Canarias (IAC). 

This research has made use of NASA's Astrophysics Data System, the SIMBAD data base, operated at CDS, Strasbourg, France, the Exoplanet Orbit Database and the Exoplanet Data Explorer at exoplanets.org, and the Exoplanets Encyclopaedia at exoplanet.eu. We also used Astropy, a community-developed core Python package for Astronomy (Astropy Collaboration et al. 2013).

This work has made use of data from the European Space Agency (ESA) mission
{\it Gaia} (https://www.cosmos.esa.int/gaia), processed by the {\it Gaia}
Data Processing and Analysis Consortium (DPAC,\\
https://www.cosmos.esa.int/web/gaia/dpac/consortium). Funding for the DPAC
has been provided by national institutions, in particular the institutions
participating in the {\it Gaia} Multilateral Agreement.

D.G. gratefully acknowledges the financial support of the \emph{Programma Giovani Ricercatori -- Rita Levi Montalcini -- Rientro dei Cervelli} (2012) awarded by the Italian Ministry of Education, Universities and Research (MIUR). Sz.Cs. thanks the Hungarian OTKA Grant K113117.
M.E. and W.D.C. were supported by NASA grant NNX16AJ11G to The University of Texas at Austin. I.R. acknowledges support from the Spanish Ministry of Economy and Competitiveness (MINECO) and the Fondo Europeo de Desarrollo Regional (FEDER) through grant ESP2016-80435-C2-1-R, as well as the support of the Generalitat de Catalunya/CERCA programme. T.H. was supported by JSPS KAKENHI Grant Number JP16K17660. M.F. and C.M.P. gratefully acknowledge the support of the
Swedish National Space Board. This work is partly supported by JSPS KAKENHI Grant Number JP18H01265.

Finally, we acknowledge our anonymous referee for a thorough review, which helped to improve the paper.
}

\section{Appendix: Issues arising from the poorly-sampled light curve of \ceight}
\label{sec:appendix}
As we mentioned in Section~5.4, and can be clearly seen in Fig.~6, the \ktwo light curve of \ceight is poorly-sampled in orbital phase. This is a result of the near-commensurability of the orbital period and the observational cadence. In this particular case, this leads to difficulty in determining the transit duration, and the physical parameters dependent on this.

The transit depth (and hence planet-to-star radius ratio) is well constrained by the data; there exists photometry close to the transit midpoint, and there is no problem in determining the out-of-transit baseline. However, there is no data covering any of the four contact points (the beginning and end of the ingress and egress phases). This results in little constraint on the duration of both the transit and of ingress and egress.

After fitting for the limb-darkening parameters as usual, we tried fixing them to the theoretical values of Sing (2010). We took the values corresponding to \logg = 4.5, \feh = 0.1, and \teff = 4500~K. We allow these values to vary slightly ($\pm 0.01$), to account for the uncertainty in the stellar parameters, and for the fact that the coefficients are tabulated only for certain values of \logg, \feh, and \teff. The allowed variation encompasses the limb-darkening parameters tabulated for neighbouring values of these parameters.

Our fits resulted in two families of solutions, revealing a degeneracy between $a/R_{\rm *}$, $b$, and the limb-darkening coefficients. The two groups of solutions result in light curve fits which look nearly identical, but which have significantly different values of $a/R_{\rm *}$, resulting in drastically different stellar densities. Instead of $a/R_{\rm *} \approx 13.8$ and $b \approx 0.2$, the second solution has $a/R_{\rm *} \approx 10.7$, $b \approx 0.65$, and limb-darkening coefficients that lie far from any tabulated values ($u_+ = 1.7$, $u_- = -0.3$). The resulting stellar density ($\approx 1400$~kg m$^{-3}$) is inconsistent with our various stellar analyses (Section~3.2). Furthermore, adopting this less-dense value results in the star lying in a region of parameter space not covered by any of the Dartmouth isochrones (Dotter et al. 2008).

We find that even when constraining the limb-darkening coefficients, a small fraction of the MCMC posterior distribution lies in a distinct region of parameter space, with a stellar density far too low to be compatible with our knowledge of the star (Fig.~12). We therefore opt both to constrain the limb-darkening coefficients and to exclude solutions with $a/R_{\rm *} < 12$ from the posterior distribution. This is illustrated in Fig.~12.

We note that previous studies have recommended fitting, rather than fixing limb-darkening coefficients, in order to avoid biasing the determination of the system parameters (Csizmadia et al. 2013; Espinoza \& Jord\'an 2015). These studies did not consider poorly-sampled light curves such as that of \ceight, however. Fortunately, \ceight lies in a region of parameter space where there is minimal difference between various tabulated limb-darkening coefficients; this is not true for all spectral types (Fig.~1 of Csizmadia et al. 2013).

\begin{figure}
	\centering
	\includegraphics[width=12cm]{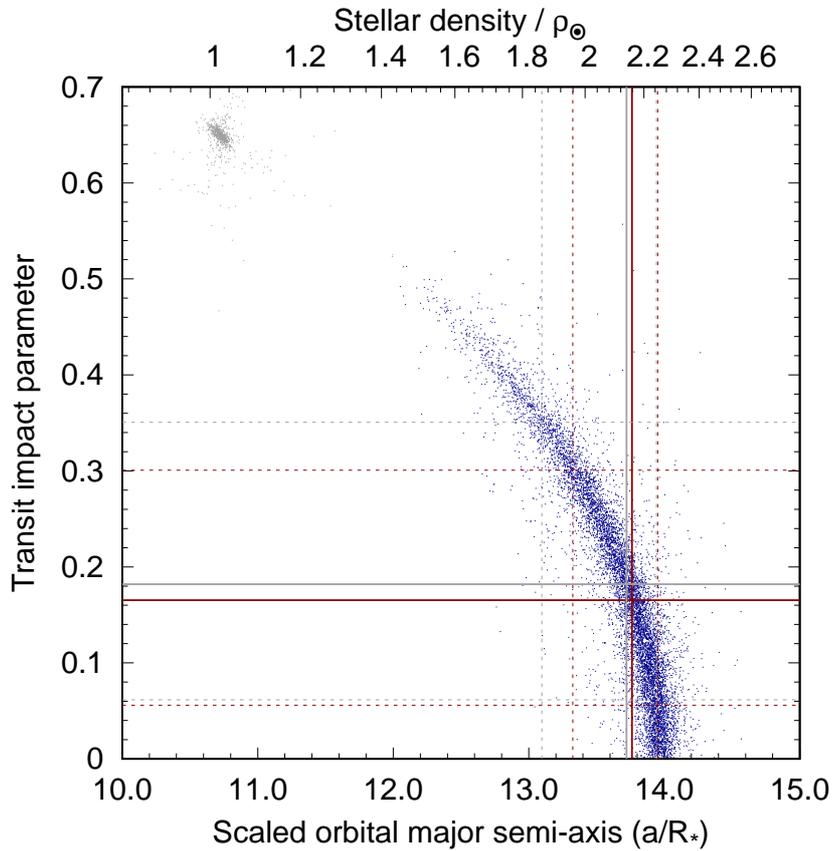}
    \caption{Posterior distribution of impact parameter and $a/R_*$, when the limb-darkening coefficients are constrained as described in the text. The corresponding stellar density is indicated along the top of the plot. A total of 10\,000 randomly-selected samples from the posterior distribution are shown, with excluded points coloured grey. The red solid line indicates our adopted solution (median of remaining points), and the red dashed lines the $1\sigma$ confidence interval. The solid and dashed grey lines indicate the solution obtained without excluding the grey points in the top left.}
    \label{fig:c8_appendix}
\end{figure}



\begin{references}


\refitem{Anderson, D. R., et al.}{2012}{MNRAS}{422}{1988}
\refitem{Andrae, R., et al.}{2018}{A\&A}{616}{A8}
\refitem{Astropy Collaboration et al.}{2013}{A\&A}{558}{A33}
\refitem{Bakos, G. \'A., L\'az\'ar, J., Papp I., S\'ari P., and Green E. M.}{2002}{PASP}{114}{974}
\refitem{Baranne, A., et al.}{1996}{A\&AS}{119}{373}
\refitem{Barnes, S. A.}{2010}{ApJ}{722}{222}
\refitem{Barnes, S. A. \& Kim, Y.-C.}{2010}{ApJ}{721}{675}
\refitem{Brown, D. J. A.}{2014}{MNRAS}{442}{1844}
\refitem{Bruntt, H., et al.}{2010}{MNRAS}{405}{1907}
\refitem{Cabrera, J., Csizmadia, S., Erikson, A., Rauer, H., and Kirste, S.}{2012}{A\&A}{548}{A44}
\refitem{Cabrera, J., et al.}{2017}{A\&A}{606}{A75}
\refitem{Csizmadia, S., et al.}{2013}{A\&A}{549}{A9}
\refitem{Csizmadia, S., et al.}{2015}{A\&A}{584}{A13}
\refitem{Dotter, A., et al.}{2008}{ApJS}{178}{89}
\refitem{Doyle, A. P., Davies, G. R., Smalley, B., Chaplin, W. J., and Elsworth, Y.}{2014}{MNRAS}{444}{3592}
\refitem{Endl, M., and Cochran, W. D.}{2016}{PASP}{128}{094502}
\refitem{Espinoza, N., and Jord\'an, A.}{2015}{MNRAS}{450}{1879}
\refitem{Frandsen, S., and Lindberg, B.}{1999} {Astrophysics with the NOT [eds. Karttunen \& Piirola]}{~}{p.~71}
\refitem{Gaia Collaboration et al.}{2016}{A\&A}{595}{A1}
\refitem{Gaia Collaboration et al.}{2018}{A\&A}{616}{A1}
\refitem{Gandolfi, D., et al.}{2015}{A\&A}{576}{A11}
\refitem{Grziwa, S., and P\"atzold, M.}{2016}{arXiv}{~}{1607.08417}
\refitem{Grziwa, S., P\"atzold, M., and Carone, L.}{2012}{MNRAS}{420}{1045}
\refitem{G\"unther, M. N., et al.}{2018}{MNRAS}{478}{4720}
\refitem{Hartman, J. D., et al.}{2015}{AJ}{149}{166}
\refitem{Hayano, Y., et al.}{2010}{Adaptive Optics Systems II.}{~}{p. 77360N}
\refitem{Hellier, C., et al.}{2015}{AJ}{150}{18}
\refitem{Hirano, T., et al.}{2016}{ApJ}{820}{41}
\refitem{Howard, A. W., et al.}{2012}{ApJS}{201}{15}
\refitem{Howell S. B., et al.}{2014}{PASP}{126}{398}
\refitem{Huber D., et al.}{2016}{ApJS}{224}{2}
\refitem{Jensen, A. G., et al.}{2012}{ApJ}{751}{86}
\refitem{J\'odar, E., et al.}{2013}{MNRAS}{429}{859}
\refitem{Kobayashi, N., et al.}{2000}{Proc. SPIE}{4008}{1056}
\refitem{Kurucz, R. L.}{2013}{Astrophysics Source Code Library}{~}{ascl:1303.024}
\refitem{Labadie, L., et al.}{2010}{Ground-based and Airborne Instrumentation for Astronomy III}{~}{p. 77350}
\refitem{Lopez, E. D., Fortney, J. J., and Miller, N.}{2012}{ApJ}{761}{59}
\refitem{Luger, R., et al.}{2016}{AJ}{152}{100}
\refitem{Mandel, K., and Agol, E.}{2002}{ApJ}{580}{L171}
\refitem{Maxted, P. F. L., Serenelli, A. M., and Southworth, J.}{2015}{A\&A}{577}{A90}
\refitem{Mayor, M., et al.}{2003}{The Messenger}{114}{20}
\refitem{Narita, N., et al.}{2015}{Journal of Astronomical Telescopes, Instruments, and Systems}{1}{045001}
\refitem{Ofir, A.}{2014}{A\&A}{561}{A138}
\refitem{Oscoz, A., et al.}{2008}{Ground-based and Airborne Instrumentation for Astronomy II}{~}{p. 701447}
\refitem{Pecault, M. J., and Mamajek, E. E.}{2013}{ApJS}{208}{9}
\refitem{Pepe, F., et al.}{2002}{A\&A}{388}{632}
\refitem{Petigura, E. A., et al.}{2018}{AJ}{155}{21}
\refitem{Pollacco, D. L., et al.}{2006}{PASP}{118}{1407}
\refitem{Queloz, D., et al.}{2001}{A\&A}{379}{279}
\refitem{Ryabchikova, T. A., Pakhomov, Y. V., and Piskunov, N. E.}{2011}{Kazan Izdatel Kazanskogo Universiteta}{153}{61}
\refitem{Ryabchikova, T., et al.}{2015}{Phys. Scr.}{90}{054005}
\refitem{Sato, B., et al.}{2005}{ApJ}{633}{465}
\refitem{Seager, S., and Deming, D.}{2010}{ARA\&A}{48}{631}
\refitem{Sestovic, M., Demory, B.-O., and Queloz, D.}{2018}{A\&A}{616}{A76}
\refitem{Sing, D. K.}{2010}{A\&A}{510}{A21}
\refitem{Sing, D. K., et al.}{2016}{Nature}{529}{59}
\refitem{Smith, A. M. S., et al.}{2014}{A\&A}{570}{A64}
\refitem{Smith, A. M. S., et al.}{2017}{MNRAS}{464}{2708}
\refitem{Soto, M. G., et al.}{2018}{MNRAS}{478}{5356}
\refitem{Southworth, J.}{2009}{MNRAS}{394}{272}
\refitem{Southworth, J.}{2011}{MNRAS}{417}{2166}
\refitem{Spake, J. J., et al.}{2018}{Nature}{557}{68}
\refitem{Spiegel, D. S., Fortney, J. J., and Sotin, C.}{2014}{PNAS}{111}{12622}
\refitem{Stassun, K. G., and Torres, G.}{2018}{ApJ}{862}{1}
\refitem{Telting, J. H., et al.}{2014}{Astron. Nachr.}{335}{41}
\refitem{Tull, R. G., MacQueen, P. J., Sneden, C., and Lambert, D. L.}{1995}{PASP}{107}{251}
\refitem{Valencia, D., Ikoma, M., Guillot, T., and Nettelmann, N.}{2010}{A\&A}{516}{A20}
\refitem{Valenti, J. A., and Fischer, D. A.}{2005}{ApJS}{159}{141}
\refitem{Valenti, J. A., and Piskunov, N.}{1996}{A\S}{118}{595}
\refitem{Vanderburg, A., and Johnson, J. A.}{2014}{PASP}{126}{948}
\refitem{Vidal-Madjar, A., et al.}{2003}{Nature}{422}{143}
\refitem{Yee, S. W., Petigura, E. A., and von Braun, K.}{2017}{ApJ}{836}{77}

\end{references}
\end{document}